\documentclass[twocolumn]{jpsj2}
\def\runauthor{Hideo {\sc Yoshioka}}
\def\d{{\rm d}}

\def\e{{\rm e}}

\def\kf{k_{\rm F}}

\def\ggs{\buildrel\textstyle > \over {\hbox{\raise0.2ex\hbox{$\sim$}}}}
\def\lls{\buildrel\textstyle < \over {\hbox{\raise0.2ex\hbox{$\sim$}}}}
\def\gsim{\,\lower0.75ex\hbox{$\ggs$}\,}
\def\lsim{\,\lower0.75ex\hbox{$\lls$}\,}

\def\tt{\tilde{t}}
\def\im{{\rm i}}
\def\ie{{\it i.e.}, }

\def\on{\omega_n}
\def\en{\epsilon_n}
\def\jo #1#2#3#4{#1 {\bf #2} (#3) #4}   

\def\PRB{Phys.\ Rev.\ B}
\def\PRL{Phys.\ Rev.\ Lett.}

\def\JPIVF{J.\ Phys.\ IV\ France}

\def\JPSJ{J.\ Phys.\ Soc.\ Jpn.}

\def\JPCS{J.\ Phys.\ Chem.\ Solids}
\def\CR{Chem.\ Rev.}
\title
{
Knight Shift and Nuclear Spin Relaxation Rate  
in a Charge-Ordered State of the One-Dimensional Extended Hubbard Model
at Quarter Filling
}

\author{
Hideo {\sc Yoshioka}\thanks{E-mail: h-yoshi@cc.nara-wu.ac.jp} 
}

\inst{
Department of Physics, Nara Women's University, Nara 630-8506
}

\recdate{\today}

\abst{
We investigate Knight shift and nuclear spin relaxation rate
in a charge ordered state of the one-dimensional extended Hubbard
model with a quarter filled band by using RPA around the mean-field
solution.  
It is shown that both quantities show splitting below the critical
temperature of the charge order, as is experimentally observed.  
The relationship between the mount of the splitting in the  both quantities and 
the charge disproportionation rate is discussed.   
}

\kword{extended Hubbard model, quarter filling, charge order, 
Knight shift, nuclear spin relaxation rate
}

\begin{document}
\maketitle

\section{Introduction}
Organic conductors have been known to exhibit a variety of states 
depending on the temperature, pressure and so on. 
Such a property is indeed due to the low dimensional electronic structure and
the mutual interaction. 
Thus,  the materials are one of the most suitable systems for 
studying exotic states realized by the interaction between electrons 
in low dimensional systems.\cite{review} 

Recently, the remarkable state with inhomogeneous but regular charge
arrangement called as the charge ordering (CO) has been found 
in the organic materials with 2:1 composition ratio $A_2 B$, 
such as (DI-DCNQI)$_2$Ag,\cite{Hiraki1998PRL,Kanoda1999JPhys,Nogami1999JPhys} (TMTTF)$_2 X$\cite{Chow2000PRL,Monceau2001PRL,Zamborszky2002PRB,Nakamura2003JPSJ} and (BEDT-TTF)$_2X$\cite{Miyagawa2000PRB,Tajima2000PRB,Chiba2001JPCS,Takano2001JPCS,Yamamoto2002PRB,W2003PRB}
with $X$ being a monovalent anion.   
Since the molecule $B$ is ionized as $B^+$ or $B^-$ in order to form a closed shell,
the average valence of $A$ becomes $-1/2$ or $+1/2$. 
As a result, the energy band composed of LUMO or HOMO of the molecule $A$ 
becomes quarter filled in terms of electrons or holes, respectively.        
Thus, the CO phenomenon is seen in several molecular conductors with the
quarter-filled band. 
Theoretically, the CO phenomenon has been analyzed based on the extended
Hubbard model with the parameters reflecting the actual band and crystal
structure, and it has been found that the phenomenon is due to the strong mutual
interaction between electrons, especially the repulsion between the
nearest neighbor molecules.\cite{Seo97JPSJ,Seo00JPSJ,Seo04CR}   

The nuclear magnetic resonance (NMR) is the powerful experimental technique which 
can explore the local electronic state. 
It has been considered to be one of the experimental evidence of the CO 
that the shift of the resonance frequency shows splitting and/or
the relaxation rate of the nuclear spin becomes multi-components. 
It is because the equivalent molecules in the absence of the CO becomes
inequivalent under the CO.     
However, to the best of our knowledge, no one show theoretically
that the Knight shift exhibit separation and the NMR relaxation rate
has several components in the presence of CO.  
It has been found that CO does not always lead to the splitting of
the Knight shift and the multi component relaxation rate. 
In the case of the extended Hubbard model with half filling, 
the Knight shift show no splitting and 
the relation rate is only one component even in the presence of CO.\cite{Yoshioka02JPSJ} 
This result is due to the fact that the particle-hole transformation,
under which the Hamiltonian is invariant, 
exchanges the charge rich site with the poor site, vise versa. 
In the state with CO at the quarter filling, 
there is no such a symmetry.     

In the present paper, 
we investigate the Knight shift and the nuclear spin relaxation rate of the charge
ordered state in the system with the quarter filled band. 
We use the one-dimensional extended Hubbard model and apply the random
phase approximation (RPA) around the mean-filed solution to the
Hamiltonian. 
It has been shown that the splitting of the both quantities occurs 
below the critical temperature of the charge ordering, $T_{\rm CO}$.  
The result is different from the case of the half-filling. 
In addition, the relationship between the amount of the splitting of the
both quantities and the charge disproportionation is discussed.  

The paper is organized as follows. 
The model we consider and it's solution by the mean-field
treatment are shown in \S.2. 
The formulation for calculating the Knight shift and the nuclear spin
relaxation rate based on the path integral method is also given in
\S.2. 
The obtained results are shown in \S.3. 
\S.4 is devoted to Summary.      

\section{Model and Formulation}
\begin{figure}
\centerline{\includegraphics[width=7.0truecm]{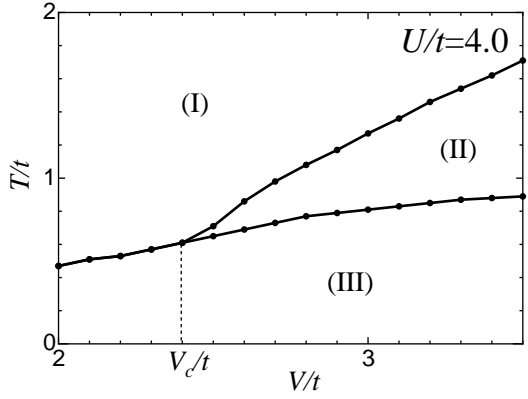}}
\caption{
The phase diagram on the plane of $V/t$ and $T/t$ at $U/t=4.0$. 
Here, (I), (II) and (III) denote the normal state,
$4\kf$-CDW and the coexistent state of $4\kf$-CDW and $2\kf$-SDW,
respectively.
}
\label{fig:phase}
\end{figure}

\begin{figure}
\centerline{\includegraphics[width=7.0truecm]{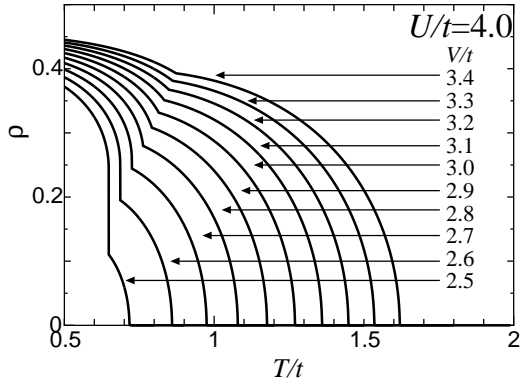}}
\caption{
The charge disproportionation $\rho$ as a function of temperature 
for the several choices of $V/t$ at $U/t = 4.0$. 
}
\label{fig:rho}
\end{figure}

We consider the one-dimensional extended Hubbard model at 
the quarter filling written by the following Hamiltonian, 
${\cal H} = {\cal H}_{\rm k} + {\cal H}_{\rm int}$, 
\begin{align}
{\cal H}_{\rm k}
=& - t \sum_{j,s} 
   \left( c_{j,s}^\dagger  c_{j+1,s}  + {\rm h.c.}\right),  
\label{eqn:Hk} \\
{\cal H}_{\rm int} =& \frac{U}{2} \sum_{j,s} 
n_{j,s} n_{j, -s} + 
V \sum_{j,s,s'} n_{j,s} n_{j+1,s'},    
\label{eqn:Hint}  
\end{align}
where $t$, $U$ and $V$ are the transfer energy between the
nearest-neighbor site, the on-site repulsion and the interaction between 
the nearest-neighbor site, respectively.   
Here,   
$c_{j,s}^\dagger$ denotes the creation
operator of the electron at the $j$-th site with spin $s = \pm$, 
$n_{j,s}=c_{j,s}^\dagger c_{j,s}$ and $n_j = \sum_s n_{j,s}$.  
The possible ordered states of the model obtained by the standard mean-filed
  approximation  
are $4\kf$-CDW (CO),
  $2\kf$-SDW and the coexistence of both states. 
The phase diagram at the finite temperature 
on the $V/t$-$T/t$ plane for $U/t=4.0$ is shown in
Fig.\ref{fig:phase},\cite{Tomio01JPCS}   where (I), (II) and (III) denote the normal state,
$4\kf$-CDW and the coexistent state of $4\kf$-CDW and $2\kf$-SDW,
respectively. 
Note that the pure $2\kf$-SDW state appears for the small $V$ region, 
which is not shown in Fig.\ref{fig:phase}.
With decreasing temperature, the transition from the normal state to the
coexistent state occurs for $V<V_c$, whereas the pure $4\kf$-CDW is
realized between them for $V>V_c$.  
Figure \ref{fig:rho} shows the charge disproportionation $\rho$ at $U/t=4.0$ 
as a function of
$T/t$ for the several choices of $V$, where the quantity $\rho$ is
defined as $\langle n_j \rangle = 1/2 - (-1)^j \rho$. 
At the transition temperature of the coexistent state, 
the charge disproportionation shows sudden increase.   
The transition from the normal state to the pure CO is the second-order. 
For $V \gg V_c$, the transition to the coexistent state is continuous, 
whereas the transition become the first order for $V \sim V_c$.\cite{Tomio01JPCS}   
In the following, we assume that 
the pure $4\kf$-CDW state (II) corresponds to the charge ordered state
without magnetic order 
realized from 5.5 K to 220 K in (DI-DCNQI)$_2$Ag and 
investigate Knight shift and the nuclear spin
relaxation rate in the state (II) 
by using RPA around the mean-field solution. 

The fluctuation around the mean-filed solution can be taken into account
by the path integral method. 
The effective action up to the second order of the charge and spin
fluctuation, which correspond to the RPA approximation where the
coupling between the different modes is neglected, 
is divided into the spin and
the charge part ( see Appendix ). 
The spin part, $S_{\sigma}$ is written as follows,  
\begin{align}
S_\sigma 
=& \sum_{0 \leq q < Q_0} \sum_{\omega_n} 
(\delta \sigma (-q, -\im \omega_n ), \delta \sigma (-q + Q_0, -\im \omega_n))
 \nonumber \\
&\times \left(
\begin{array}{cc}
 A_\sigma(q, \im \omega_n)& B_\sigma(q, \im \omega_n) \\
 B_\sigma(q, \im \omega_n)& A_\sigma(q-Q_0, \im \omega_n)
\end{array}
\right) \nonumber \\
&\times \left(
\begin{array}{c}
\delta \sigma (q, \im \omega_n )  \\
\delta \sigma (q-Q_0, \im \omega_n )  
\end{array}
\right),
\label{eqn:s-action}
\end{align}
where $Q_0 = \pi/a$ with $a$ being the lattice spacing and 
\begin{align}
 A_\sigma (q, \im \omega_n) &= - \frac{v_s(q)}{N_L} \left\{ 1 + 4 v_s(q) K(q, \im
 \omega_n) \right\},
\label{eqn:AS}
\\
 B_\sigma (q, \im \omega_n) &= - \frac{4 v_s(q) v_s(q-Q_0)}{N_L} H(q, \im
 \omega_n), 
\label{eqn:BS}
\end{align}
with $v_s (q) = - U/4$ and  $N_L$ is the total number of the lattice. 
Here, the quantities $K(q,\im \omega_n)$ and $H(q,\im \omega_n)$
are defined as follows, 
\begin{align}
& K(q,\im \omega_n ) \nonumber \\
=& - \frac{1}{N_L} \sum_k 
\Bigg\{
\frac{1}{4} \left( 1 +
 \frac{\epsilon_k}{E_k}\frac{\epsilon_{k+q}}{E_{k+q}} +
 \frac{\Delta}{E_k}\frac{\Delta}{E_{k+q}} \right) \nonumber \\
&\times 
\Big(
\frac{f(E_k - \mu) - f(E_{k+q} - \mu)}{\im \omega_n + E_k - E_{k+q} }
 \nonumber \\
&\hspace{4em}+
\frac{f(-E_k - \mu) - f(-E_{k+q} - \mu)}{\im \omega_n - E_k + E_{k+q} }
\Big) \nonumber \\
&+ \frac{1}{4} \left( 1 -
 \frac{\epsilon_k}{E_k}\frac{\epsilon_{k+q}}{E_{k+q}} -  
 \frac{\Delta}{E_k}\frac{\Delta}{E_{k+q}} \right) \nonumber \\
&\times 
\Big(
\frac{f(E_k - \mu) - f(-E_{k+q} - \mu)}{\im \omega_n + E_k + E_{k+q} }
 \nonumber \\
&\hspace{4em}+
\frac{f(-E_k - \mu) - f(E_{k+q} - \mu)}{\im \omega_n - E_k - E_{k+q} }
\Big)
\Bigg\}, \label{eqn:K}
\end{align}
and 
\begin{align}
& H(q,\im \omega_n ) \nonumber \\
=& - \frac{1}{N_L} \sum_k 
\Bigg\{
\frac{1}{4} \left( 
 \frac{\Delta}{E_k} + \frac{\Delta}{E_{k+q}} \right) \nonumber \\
&\times 
\Big(
\frac{f(E_k - \mu) - f(E_{k+q} - \mu)}{\im \omega_n + E_k - E_{k+q} }
 \nonumber \\
&\hspace{4em}-
\frac{f(-E_k - \mu) - f(-E_{k+q} - \mu)}{\im \omega_n - E_k + E_{k+q} }
\Big) \nonumber \\
&+ \frac{1}{4} \left( 
 \frac{\Delta}{E_k}- \frac{\Delta}{E_{k+q}} \right) \nonumber \\
&\times 
\Big(
\frac{f(E_k - \mu) - f(-E_{k+q} - \mu)}{\im \omega_n + E_k + E_{k+q} }
 \nonumber \\
&\hspace{4em}-
\frac{f(-E_k - \mu) - f(E_{k+q} - \mu)}{\im \omega_n - E_k - E_{k+q} }
\Big)
\Bigg\} \nonumber \\ 
=& H(q-Q_0, \im \omega_n).
\label{eqn:H}
\end{align}
Here, $\mu$ is the chemical, 
$\epsilon_k = - 2 t \cos ka$ and $E_k = \sqrt{\epsilon_k^2 +
\Delta^2}$.  
The energy gap $\Delta \geq 0$ which opens at $\pm \pi/(2a)$ is related to the charge
disproportionation $\rho$ as 
\begin{align}
 \Delta = - 2 v_c (Q_0) \rho.
\label{eqn:DvsRho}  
\end{align}
where $v_c(q) = U/4 + V \cos qa$. 
By using the action, 
the correlations of the spin fluctuation are obtained as follows, 
\begin{align}
 &\langle \delta \sigma (q, \im \omega_n) \delta \sigma (-q,-\im
 \omega_n) \rangle \nonumber \\
=& 
\frac{1}{2} \frac{A_\sigma (q-Q_0, \im \omega_n)}
{A_\sigma (q, \im \omega_n) A_\sigma (q-Q_0, \im \omega_n) -
 B_\sigma^2(q, \im \omega_n)}, \label{eqn:COL0}\\
 &\langle \delta \sigma (q, \im \omega_n) \delta \sigma (-q+Q_0,-\im
 \omega_n) \rangle \nonumber \\
=& 
- \frac{1}{2} \frac{B_\sigma (q, \im \omega_n)}
{A_\sigma (q, \im \omega_n) A_\sigma (q-Q_0, \im \omega_n) -
 B_\sigma^2(q, \im \omega_n)},
\label{eqn:COL1}
\end{align}
for $-\pi/a \leq q < \pi/a$ 
and the others are zero.
Note that 
the CO leads to the coupling between $q$-component and
$-q+Q_0$-component in the fluctuation because the spacing between the
equivalent site becomes twice. 
In the case of the half filling, \ie $\mu = 0$, 
such a coupling diminishes because 
the quantity $H(q,\on)$ in eq. (\ref{eqn:H}) vanishes. 
 
The spin susceptibility in the site representation,  $\tilde{\chi}_\sigma (x_i, x_j ; \im \omega_n)$
is obtained as follows, 
\begin{align}
 &\tilde{\chi}_\sigma (x_i, x_j ; \im \omega_n) \nonumber \\
=& \frac{1}{N_L^2} \sum_{q,q'} \e^{-\im (q'x_i + qx_j)} 
\int_0^\beta {\rm d} \tau \e^{\im \omega_n \tau} 
\langle T_\tau m(q',\tau) m(q,0)
 \rangle \nonumber \\
=& \frac{1}{N_L^2} \sum_{q,q'}
 \e^{-\im (q'x_i + qx_j)} \nonumber \\
&\times \left\{
\langle \delta \sigma (q, \im \omega_n) \delta \sigma (q', -\im
 \omega_n) \rangle + \delta_{q',-q} \frac{1}{2} \frac{N_L}{v_s(q)}
\right\} \nonumber \\
%
%
\equiv& \frac{1}{N_L} \sum_q \e^{\im q (x_i - x_j)} \nonumber \\
\times&\left\{
\chi_\sigma (q,-q;\im \omega_n) + (-1)^i \chi_\sigma (q,-q+Q_0;\im \omega_n)
\right\},   
\end{align}
where 
\begin{align}
 & \chi_\sigma (q,-q;\im \omega_n) \nonumber \\ =& 
\frac{\langle \delta \sigma (q, \im \omega_n) \delta \sigma (-q, -\im
 \omega_n) \rangle}{N_L} + \frac{1}{2v_s(q)}, \label{eqn:CHI-0}\\
 & \chi_\sigma (q,-q+Q_0;\im \omega_n) \nonumber \\ =& 
\frac{\langle \delta \sigma (q, \im \omega_n) \delta \sigma (-q+Q_0, -\im
 \omega_n) \rangle}{N_L}. 
\label{eqn:CHI-1}
\end{align}
The Knight shift $S_i$ at the location $x_i$ is proportional to the local
magnetic moment at the $i$-th site under
the uniform and static magnetic magnetic field. 
Therefore, $S_i$ is written by using eqs.(\ref{eqn:CHI-0}) and (\ref{eqn:CHI-1}) as follows,
\begin{align}
 S_i &\propto \sum_{j} \tilde{\chi}_\sigma (x_i,x_j;0) \nonumber \\
&\propto \chi_\sigma (0,0;0) + (-1)^i \chi_\sigma (0,Q_0;0).  
\end{align}
Thus, the average between  the Knight shift at the charge rich site and
that at the poor site is proportional to the magnetic
susceptibility\cite{Hiraki1998PRL}. 
When the hyperfine coupling constant is assumed to be independent of the
wave number, the nuclear spin relaxation rate $R_i = (T_1 T)_i^{-1}$ is
written as follows, 
\begin{align}
 R_i \propto& \lim_{\omega \to 0} 
\frac{{\rm Im} \tilde{\chi}_\sigma (x_i,x_i;\omega + \im
 \delta)}{\omega} \nonumber \\
=& \frac{1}{N_L} \sum_q \lim_{\omega \to 0}
\Bigg\{
\frac{{\rm Im} \chi_\sigma (q,-q;\omega + \im
 \delta)}{\omega} \nonumber \\
& + (-1)^i
\frac{{\rm Im} \chi_\sigma (q,-q+Q_0;\omega + \im
 \delta)}{\omega}
\Bigg\}, 
\end{align} 
where $\delta = +0$. 
Note that the transverse susceptibility is the same as the longitudinal
one because the order of the spin does not appear in the region (II).   
The charge rich and poor sites are the odd site ($i$=odd) and the even
site ($i$=even), respectively. 
In the case of the half-filling, \ie $\mu = 0$,  
the quantity, $H (q, \im \omega_n)$ in eq. (\ref{eqn:H}) vanishes, as is
already noted.  
Therefore,  
the Knight shift and the relaxation rate do not show splitting. 
However, in the present quarter-filling case, 
the splitting of both quantities appear, as will be shown in the next
section concretely.   

\section{Results}

In this section, we discuss the Knight shift and the nuclear spin
relaxation rate in the pure CO state (II) in the phase diagram of
Fig. \ref{fig:phase}.  

\begin{figure}[hob]
\centerline{\includegraphics[width=7.0true cm]{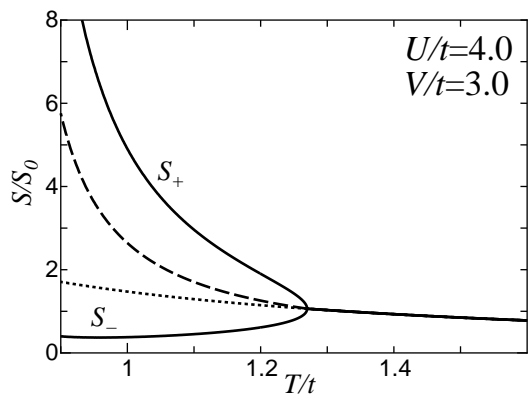}}
\caption{
The normalized Knight shift at the charge-rich site $S_+/S_0$ and that at the poor site
 $S_-/S_0$ as a function of the temperature for the pure charge ordered
 state in the case of  $U/t=4.0$ and $V/t=3.0$ 
Here $S_0$ is the shift of the non-interaction case at $T=0$. 
The dashed curve expresses the 
the average of $S_+$ and $S_-$, 
\ie magnetic susceptibility. 
The shift without taking account of the charge order is denoted by 
the dotted curve. 
}
\label{fig:shift1}
\end{figure}

Figure \ref{fig:shift1} shows the Knight shift at the charge rich site $S_+$ and that at
the poor site $S_-$ as a function of temperature for $U/t=4.0$ and
$V/t=3.0$. 
These quantities are normalized by the shift in the non-interacting case at $T=0$,
$S_0$. 
The shift splits into the two components below $T_{\rm CO}$ 
and the quantity at the rich site is larger than that at the poor site. 
The fact can be easily understood by considering the case where the
charge order is complete, 
\ie $\rho = 1/2$. 
In the charge poor site where there are no carriers, 
the spin excitation is impossible. 
Therefore the shift is equal to zero. 
On the other hand, in the charge rich site with one electron per site, 
the spin excitation is possible and the Knight shift is finite. 
The dashed curve and dotted one express the average of $S_+$ and $S_-$,
\ie the uniform and static magnetic susceptibility and the quantity 
in the absence of the charge order, respectively.  
The magnetic susceptibility is enhanced by the appearance of the charge
order. 
It is due to the enhancement of the density of state at the Fermi energy 
by opening  the gap at $\pm 2 \kf$. 
It should be noticed that 
the magnetic susceptibility does not show any singularity at the
critical temperature of the charge order,\cite{Tomio2002DT} which has been observed experimentally.\cite{Hiraki1998PRL}    

\begin{figure}[htb]
\centerline{\includegraphics[width=7.0truecm]{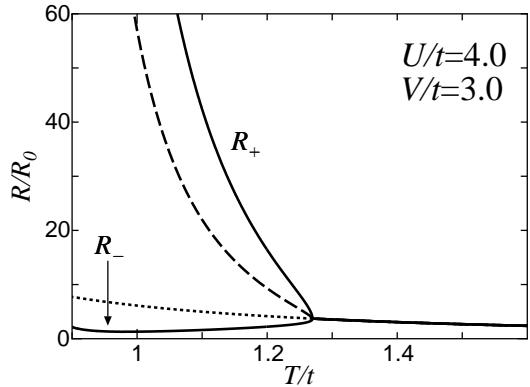}}
\caption{
The normalized nuclear spin relaxation rate at the charge-rich site $R_+/R_0$ and that at the poor site
 $R_-/R_0$ as a function of the temperature for the pure charge ordered
 state in the case of  $U/t=4.0$ and $V/t=3.0$ 
Here $R_0$ is the rate of the non-interaction case at $T=0$. 
The dashed curve expresses the 
the average of $R_+$ and $R_-$. 
The rate without taking account of the charge order is denoted by 
the dotted curve. 
}
\label{fig:rate1}
\end{figure}

The nuclear spin relaxation rate for $U/t=4.0$ and $V/t=3.0$ 
normalized by the quantity in the
absence of the interaction at $T=0$, $R_0$  
is shown in Fig. \ref{fig:rate1}, 
where $R_+$ and $R_-$ are $(T_1 T)^{-1}$ at the charge rich site and
that of poor site, respectively.   
Here the dashed curve expresses the average of $R_+$ and $R_-$ and 
the rate in the absence of CO is shown by the dotted curve.   
As is seen in the Knight shift, 
the rate shows splitting and  
the quantity $R_+$ becomes larger $R_-$
because the spin fluctuation at the charge rich site is larger than that
at the poor site.  
Note that the average seems to have a singularity at $T = T_{\rm CO}$,
which is different from the magnetic susceptibility. 

Next, the relationship between the mount of the splitting in the these
quantities and the charge disproportionation $\rho$ is discussed. 
The experiments have been analyzed by assuming 
that the relative shift, $S_+/(S_+ + S_-)$ or $S_-/(S_+ + S_-)$
measures the local charge\cite{Hiraki1998PRL} and that      
the relaxation rate is proportional to the local charge  
$(1/2 \pm \rho)^2$,\cite{Zamborszky2002PRB} respectively.  
If the assumptions are valid, 
the following relationship should hold, 
\begin{align}
 \frac{S_+ - S_-}{S_+ + S_+} &= 2\rho, \label{eqn:aS}\\
 \frac{R_+ - R_-}{R_+ + R_+} &= \frac{2\rho}{1/2 + 2 \rho^2}.
\label{eqn:aR}
\end{align}
Figures \ref{fig:shift2} and \ref{fig:rate2} show
the l.h.s. of eq.(\ref{eqn:aS}) and that of eq.({\ref{eqn:aR})
as a function of the r.h.s. of the respective equations.   
\begin{figure}[htb]
\centerline{\includegraphics[width=7.0truecm]{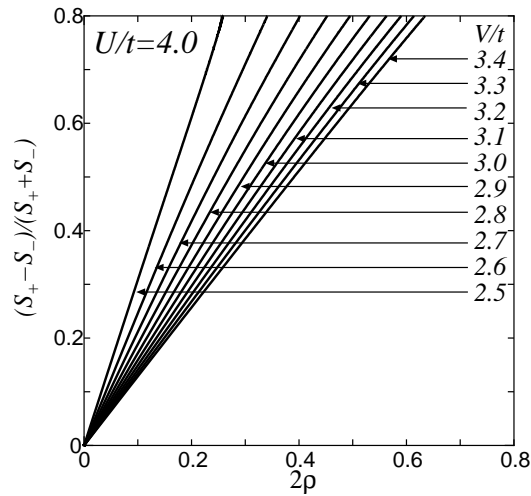}}
\caption{
The mount of splitting of the Knight shift $S_+ - S_-$ normalized by
 $S_+ + S_-$ as a function of $2\rho$ for several
 choices of $V/t$ at $U/t=4.0$.
}
\label{fig:shift2}
\end{figure}
\begin{figure}[htb]
\centerline{\includegraphics[width=7.0truecm]{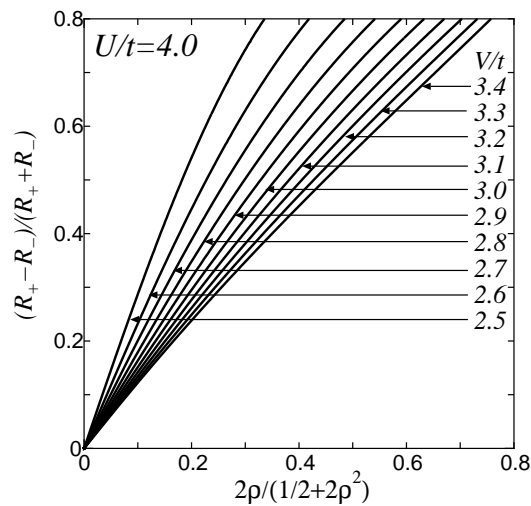}}
\caption{
The mount of splitting of the nuclear spin relaxation rate $R_+ - R_-$ normalized by
 $R_+ + R_-$ as a function of $2\rho/(1/2+2\rho^2)$ for several
 choices of $V/t$ at $U/t=4.0$.
}
\label{fig:rate2}
\end{figure}
In Fig. \ref{fig:shift2},
the relative shift is well proportional to $2 \rho$, but 
the coefficient becomes larger than unity. 
The coefficient approaches to unity when $V$ increases. 
Similarly, the quantity $(R_+ - R_-)/(R_+ + R_-)$ approaches to 
$2 \rho / (1/2 + 2 \rho^2)$ for large  $V$. 
Thus, the assumptions, eqs. (\ref{eqn:aS}) and (\ref{eqn:aR}), seems to hold
approximately well for the large $V$ in which the CO state becomes
robust,  
but with decreasing $V$ such relationship become worse.  
It should be noted that 
eqs.(\ref{eqn:aS}) and (\ref{eqn:aR}) gives the smaller value of the
charge disproportionation that the actual one.\cite{Zamborszky2002PRB} 

\section{Summary}
In the present paper, 
we investigated the Knight shift and the nuclear spin relaxation rate
in the state with charge ordering based on the one-dimensional extended Hubbard
model with a quarter filled band with taking account of the fluctuation
around the mean-filed solution by RPA.   
It is shown that both quantities shows splitting below the critical
temperature of the charge order. 
The result is different from the case of the half-filling where 
the splitting does not occur even in the CO state due to the
particle-hole symmetry. 
The quantities at the charge rich site are larger than those at the poor
site due to the larger spin fluctuation.   
Thus, in the materials with the quarter filling, 
the splitting of these quantities can be considered to be the
experimental evidence of appearance of the charge order as has been
already experimentally insisted.
The relationship between the amount of the splitting normalized by the
average is discussed. 
It is shown that    
the assumptions used in the analysis of the experiments
hold well when $V$ becomes larger, 
\ie the CO is robust, 
though the estimated mount of the charge disproportionation becomes
generally smaller compared with the actual value.

In the present study, we used the method in which  
we apply the mean-field treatment to the present model and take into
account of the fluctuation around the solution by using RPA. 
Though this method can succeed to show the splitting of the Knight shift
and the relaxation rate, 
the transport properties of the actual materials such as
(DI-DCNQI)$_2$Ag cannot be described. 
The materials show the insulating behavior even at the room temperature,
\ie $T > T_{\rm CO}$,\cite{Hiraki1998PRL}. 
In addition, the derivative of the resistivity in terms of the temperature has a
peak at $T = T_{\rm CO}$.\cite{Ito04PRL} where the Metal-Insulator
transition occurs. 
On the other hand, the present mean-field treatment leads to 
the metallic state both $T > T_{\rm CO}$ and $T < T_{\rm CO}$. 
That is, the metal insulator transition do not happen even at $T =
T_{\rm CO}$ because the energy gap opens at not $\pm \kf$ but $\pm 2
\kf$. 
Therefore the further studies are necessary to describe the CO state
consistently in both the transport and magnetic properties.

\section*{Acknowledgments}
The author would like to H. Fukuyama, Y. Suzumura, M. Ogata, K. Kanoda,
S.E. Brown, 
H. Seo, M. Tsuchiizu and Y. Fuseya  
for useful discussion.
This work was supported by Grant-in-Aid for Scientific Research on
Priority Area of Molecular Conductors (No.16038218) and 
Grant-in-Aid  for Scientific Research (C) (No.15540343)
from MEXT. 

\appendix
\section{Effective action of the fluctuation around the pure $4\kf$-CDW state} 

The partition function corresponding to the Hamiltonian, (\ref{eqn:Hk}) and
(\ref{eqn:Hint}), is written by utilizing the Stratonovich-Hubbard
transformation as follows, 
\begin{align}
 Z =& \int {\cal D}(c^*,c) {\cal D}\rho {\cal D} \sigma \e^{-(S_{\rm k} +
 S_{\rm int})}, \\
 S_{\rm k} =& \int_0^\beta \d \tau  
\sum_{k,s} c^*_{k,s}(\tau) (\partial_\tau - \mu + \epsilon_k)
 c_{ks}(\tau) ,  \\
S_{\rm int}  =& \int_0^\beta \d \tau 
\sum_q \nonumber \\
\times& \Bigg\{ \frac{v_c(q)}{N_L}\left[-\rho(q,\tau)\rho(-q,\tau) + 2
 \rho(q,\tau) n(-q,\tau) 
\right] \nonumber \\
+& \frac{v_s(q)}{N_L}\left[-\sigma(q,\tau)\sigma(-q,\tau) + 2
 \sigma(q,\tau) m(-q,\tau) 
\right]
\Bigg\},
\end{align}
where $\mu$ is the chemical potential, 
$N_L$ is the total number of sites,  
$\beta = 1/T$,  
$v_c(q) = U/4 + V \cos q a$, $v_c(q) = - U/4$,  
$n(q,\tau) = \sum_{k,s} c^*_{k+q,s}(\tau) c_{k,s}(\tau)$ and 
$m(q,\tau) = \sum_{k,s} s c^*_{k+q,s}(\tau) c_{k,s}(\tau)$.  
Here we assume that the pure $4\kf$-CDW is realized. 
Then the Stratonovich-Hubbard fields, $\rho(q,\tau)$ and
$\sigma(q,\tau)$, are rewritten as follows, 
\begin{align}
 \rho(q,\tau) &= 
\begin{cases}
 \rho(0) + \delta \rho(0,\tau), & \text{for $q=0$} \\
 \rho(Q_0) + \delta \rho(Q_0,\tau), & \text{for $q=Q_0$} \\
 \delta \rho(q,\tau), & \text{for the others} 
\end{cases} \\
 \sigma(q,\tau) &= \delta \sigma (q,\tau). 
\end{align}
When eqs. (A$\cdot$4) and (A$\cdot$5) are substituted into (A$\cdot$3),  
the action of the interaction part is divided as $S_{\rm int} =
S_{{\rm int},0} + S_{{\rm int},0}^{e} + S_{{\rm int},1} + S_{{\rm
int},2}$, where
\begin{align}
 S_{{\rm int},0} =& - \beta \frac{v_c(0)}{N_L} \rho(0)^2 - \beta
 \frac{v_c(Q_0)}{N_L} \rho(Q_0)^2, \\
 S_{{\rm int},0}^e =& \int_0^\beta \d \tau \nonumber \\ 
\times& \left\{
2 \frac{v_c(0)}{N_L} \rho(0) n(0,\tau) + 2 \frac{v_c(Q_0)}{N_L} \rho(Q_0) n(Q_0,\tau) 
\right\}, \\
S_{{\rm int},1} =& \int_0^\beta \d \tau \bigg\{
\sum_{q \neq 0, Q_0} 2 \frac{v_c(q)}{N_L} \delta \rho (q,\tau) n(-q,\tau) \nonumber \\
&+ 2 \frac{v_c(0)}{N_L} ( n(0,\tau) - \rho(0) ) \delta \rho(0,\tau)
 \nonumber \\
&+ 2 \frac{v_c(Q_0)}{N_L} ( n(Q_0,\tau) - \rho(Q_0) ) \delta
 \rho(Q_0,\tau) \nonumber \\
&+ \sum_{q} 2 \frac{v_s(q)}{N_L} \delta \sigma (q,\tau) m(-q,\tau) 
\bigg\}, \\
S_{{\rm int},2} =& - \int_0^\beta \d \tau \sum_q \bigg\{
\frac{v_c(q)}{N_L} \delta \rho (q,\tau) \delta \rho (-q,\tau) \nonumber
 \\
&+
\frac{v_s(q)}{N_L} \delta \sigma (q,\tau) \delta \sigma (-q,\tau) 
\bigg\}. 
\end{align}
By integrated out the electron degree of freedom, the partition function
is calculated as,
\begin{align}
 Z = \e^{-S_{{\rm int},0}} Z_e^0 \int {\cal D}\delta \rho {\cal D}\delta
 \sigma
\e^{-S_{{\rm int},2}} \langle \e^{-S_{{\rm int},1}} \rangle, 
\end{align} 
where $Z_e^0 = \int {\cal D}(c^*,c) \e^{-(S_{\rm k} + S_{{\rm
int},0}^e)}$ and $\langle \cdots \rangle$ denotes 
the average in terms of $S_{\rm k} + S_{{\rm int},0}^e$ given as
follows, 
\begin{align}
 & S_{\rm k} + S_{{\rm int},0}^e \nonumber \\
=& \int_0^\beta \d \tau \sum_{0<k\leq
 Q_0,s}
(c_{k,s}^*(\tau), c_{k-Q_0,s}^*(\tau)) \nonumber \\
&\times
\left(
\begin{array}{cc}
 \partial_\tau - \mu + \epsilon_k & \Delta \\
\Delta & \partial_\tau - \mu - \epsilon_k
\end{array}
\right) 
\left(
\begin{array}{c}
 c_{k,s}(\tau) \\
 c_{k-Q_0,s}(\tau)
\end{array}
\right) \nonumber \\
=& \int_0^\beta \d \tau \sum_{0<k\leq
 Q_0,s}
(c_{k,s}^*(\tau), c_{k-Q_0,s}^*(\tau)) \nonumber \\
&\times \left(- \overleftrightarrow{G}(k,\tau) \right)^{-1}
\left(
\begin{array}{c}
 c_{k,s}(\tau) \\
 c_{k-Q_0,s}(\tau)
\end{array}
\right),
\end{align}
with $\Delta = 2 v_c (Q_0) \rho(Q_0)/N_L \equiv -2 v_c(Q_0) \rho$.
Here
\begin{align}
 \overleftrightarrow{G}(k,\tau) =&
\begin{pmatrix}
 G(k,\tau) & F(k,\tau) \\
 F(k-Q_0,\tau) & G(k-Q_0,\tau)
\end{pmatrix} 
\nonumber \\
=&
\frac{1}{\beta} \sum_{\epsilon_n} \e^{-\im \epsilon_n \tau}
 \nonumber \\ 
& \times
\left(
\begin{array}{cc}
  G(k, \im \epsilon_n) & F(k, \im \epsilon_n)\\
  F(k-Q_0, \im \epsilon_n) & G(k-Q_0, \im \epsilon_n) 
\end{array}
\right),
\end{align}
where
\begin{align}
 G(k,\im \epsilon_n) &= \frac{u_k^2}{\im \epsilon_n + \mu - E_k}
+ \frac{v_k^2}{\im \epsilon_n + \mu + E_k}, \\
 F(k,\im \epsilon_n) &= \frac{u_k v_k}{\im \epsilon_n + \mu - E_k}
- \frac{u_k v_k}{\im \epsilon_n + \mu + E_k}, 
\end{align} 
for $- Q_0 < k \leq Q_0$
with $E_k = \sqrt{ \epsilon_k^2 + \Delta^2}$ ($\Delta > 0$) and 
\begin{align}
 u_k = \sqrt{\frac{1}{2}\left(1 + \frac{\epsilon_k}{E_k}\right)}, \\
 v_k = \sqrt{\frac{1}{2}\left(1 - \frac{\epsilon_k}{E_k}\right)}. 
\end{align}
We expand $\langle \e^{-S_{{\rm int},1}} \rangle$ up to the second
order.  
The condition $\langle S_{{\rm int},1} \rangle = 0$ gives rise to 
the following equations, 
\begin{align}
 \rho (0) =& \frac{N_L}{2} = 2 \sum_{0<k\leq Q_0} \left\{f(-E_k - \mu) + 
f(E_k - \mu) \right\}, \label{eqn:Ne}\\
 \Delta =& 4v_c(Q_0) \frac{1}{N_L} \sum_{0<k\leq Q_0}
\frac{\Delta}{E_k}
\left\{f(E_k - \mu) - 
f(-E_k - \mu) \right\}.
\label{eqn:gapII}
\end{align}
Note that the chemical potential $\mu$ is determined by
eq.(\ref{eqn:Ne}) and eq.(\ref{eqn:gapII}) is the self-consistent
equation of $\Delta$. 
On the other hand, 
the quantity $\langle S_{{\rm int} 1}^2 \rangle / 2$ is calculated as
follows,
\begin{align}
 & \frac{\langle S_{{\rm int} 1}^2 \rangle}{2} \nonumber \\ =& 
\frac{1}{2} \int_0^\beta \d \tau \d \tau' \sum_{q,q'} \nonumber \\
&\times\bigg\{
\frac{2 v_c(q)}{N_L}\frac{2 v_c(q')}{N_L} \delta \rho(q,\tau) \delta \rho(q',\tau')
\langle : n(-q,\tau): :n(-q',\tau'): \rangle \nonumber \\
&+ \frac{2 v_s(q)}{N_L}\frac{2 v_s(q')}{N_L} \delta \sigma(q,\tau) \sigma \rho(q',\tau')
\langle : m(-q,\tau): :m(-q',\tau'): \rangle
\bigg\} \nonumber \\
=& 4 \int_0^\beta \d \tau \d \tau' \frac{1}{N_L} \sum_{q,q'} \nonumber
 \\
&\times \left\{
v_c(q) v_c(q') \delta \rho (q,\tau) \delta \rho (q',\tau') 
+ v_s(q) v_s(q') \delta \sigma (q,\tau) \delta \sigma (q',\tau') 
\right\} \nonumber \\
&\times \left\{
\delta_{q+q',0} K (q, \tau-\tau') + \delta_{q+q',Q_0} H (q, \tau-\tau')  
\right\},
\end{align}
where $:A: = A - \langle A \rangle$ and
\begin{align}
 & K(q,\tau) \nonumber \\
=& - \frac{1}{N_L} \sum_k \bigg\{
G(k-q,-\tau) G(k,\tau) + F(k-q,-\tau) F(k,\tau)
\bigg\} \nonumber \\
=& \frac{1}{\beta} \sum_{\omega_n} \e^{-\im \omega_n \tau} K(q,\im
 \omega_n), \\
 & H(q,\tau) \nonumber \\
=& - \frac{1}{N_L} \sum_k \bigg\{
G(k-q,-\tau) F(k,\tau) + F(k-q,-\tau) G(k,\tau)
\bigg\} \nonumber \\
=& \frac{1}{\beta} \sum_{\omega_n} \e^{-\im \omega_n \tau} H(q,\im
 \omega_n).
\end{align}
The quantities, $K(q, \im \omega_n)$ and $H(q, \im \omega_n)$ are given
as follows, 
\begin{align}
 K(q,\im \on) &= - \frac{1}{\beta N_L} \sum_{k, \en} \big\{
G(k-q, \im \en) G(k, \im \en + \im \on) \nonumber \\ 
&+ F(k-q, \im \en) F(k, \im \en + \im \on) 
\big\}, \\
 H(q,\im \on) &= - \frac{1}{\beta N_L} \sum_{k, \en} \big\{
G(k-q, \im \en) F(k, \im \en + \im \on) \nonumber \\ 
&+ F(k-q, \im \en) G(k, \im \en + \im \on) 
\big\},
\end{align}
and obtained in eq.(\ref{eqn:K}) and (\ref{eqn:H}), respectively. 
When the Fourier transformation, $\delta \nu (q,\im \omega_n) =
1/\sqrt{\beta} \int_0^\beta \d \tau \exp(-\im \omega_n \tau) \delta \nu
(q,\tau)$ ($\nu = \rho, \sigma$)
is introduced, eq.(A$\cdot$19) is written as follows,
\begin{align}
 &\frac{\langle S_{{\rm int}1}^2 \rangle}{2} \nonumber \\ 
=& \sum_{0<q\leq Q_0}
 \sum_{\omega_n} \left(\delta \rho(-q,-\im \omega_n), \delta
 \rho(-q+Q_0,-\im \omega_n) \right) \nonumber \\
&\times
\begin{pmatrix}
 \dfrac{4 v_c^2(q)}{N_L} K (q, \im \omega_n) & \dfrac{4 v_c(q)v_c(q-Q_0)}{N_L} H (q, \im \omega_n) \\
\dfrac{4 v_c(q)v_c(q-Q_0)}{N_L} H (q, \im \omega_n) & \dfrac{4 v_c^2(q-Q_0)}{N_L} K (q-Q_0, \im \omega_n) 
\end{pmatrix} \nonumber \\
&\times 
\begin{pmatrix}
 \delta \rho (q,\im \omega_n) \\ \delta \rho (q-Q_0,\im \omega_n)
\end{pmatrix} \nonumber \\
+& 
 \sum_{0<q\leq Q_0}
 \sum_{\omega_n} \left(\delta \sigma(-q,-\im \omega_n), \delta
 \sigma(-q+Q_0,-\im \omega_n) \right) \nonumber \\
&\times
\begin{pmatrix}
 \dfrac{4 v_s^2(q)}{N_L} K (q, \im \omega_n) & \dfrac{4 v_s(q)v_s(q-Q_0)}{N_L} H (q, \im \omega_n) \\
\dfrac{4 v_s(q)v_s(q-Q_0)}{N_L} H (q, \im \omega_n) & \dfrac{4 v_s^2(q-Q_0)}{N_L} K (q-Q_0, \im \omega_n) 
\end{pmatrix} \nonumber \\
&\times 
\begin{pmatrix}
 \delta \sigma (q,\im \omega_n) \\ \delta \sigma (q-Q_0,\im \omega_n)
\end{pmatrix}.
\end{align} 
On the other hand, $S_{{\rm int}2}$ is written as follows,
\begin{align}
 & S_{{\rm int}2}  \nonumber \\ 
=& - \sum_{0<q\leq Q_0}
 \sum_{\omega_n} \left(\delta \rho(-q,-\im \omega_n), \delta
 \rho(-q+Q_0,-\im \omega_n) \right) \nonumber \\
&\times
\begin{pmatrix}
 \dfrac{v_c(q)}{N_L} & 0 \\
0 & \dfrac{v_c(q-Q_0)}{N_L} 
\end{pmatrix} 
\begin{pmatrix}
 \delta \rho (q,\im \omega_n) \\ \delta \rho (q-Q_0,\im \omega_n)
\end{pmatrix} \nonumber \\
&- 
 \sum_{0<q\leq Q_0}
 \sum_{\omega_n} \left(\delta \sigma(-q,-\im \omega_n), \delta
 \sigma(-q+Q_0,-\im \omega_n) \right) \nonumber \\
&\times
\begin{pmatrix}
 \dfrac{v_s(q)}{N_L} & 0 \\
0 & \dfrac{v_s(q-Q_0)}{N_L} 
\end{pmatrix} 
\begin{pmatrix}
 \delta \sigma (q,\im \omega_n) \\ \delta \sigma (q-Q_0,\im \omega_n)
\end{pmatrix}. 
\end{align}
Thus the effective action $S_{\rm eff} = S_{{\rm int}2} - \langle
S_{{\rm int}1}^2 \rangle/2$ is given by the sum of the charge and spin
part, $ \sum_{\nu = \rho, \sigma} S_\nu$, 
where 
\begin{align}
{\cal S}_\nu
=& \sum_{0 < q \leq Q_0} \sum_{\omega_n} 
(\delta \nu (-q, -\im \omega_n ), \delta \nu (-q + Q_0, -\im \omega_n))
 \nonumber \\
&\times \left(
\begin{array}{cc}
 A_\nu(q, \omega_n)& B_\nu(q, \omega_n) \\
 B_\nu(q, \omega_n)& A_\nu(q-Q_0, \omega_n)
\end{array}
\right) \nonumber \\
&\times \left(
\begin{array}{c}
\delta \nu (q, \im \omega_n )  \\
\delta \nu (q-Q_0, \im \omega_n )  
\end{array}
\right),
\end{align}
with
\begin{align}
A_\rho (q, \omega_n) &= - \frac{v_c(q)}{N_L} \left\{ 1 + 4 v_c(q) K(q, \im
 \omega_n) \right\}, \\
 B_\rho (q, \omega_n) &= - \frac{4 v_c(q) v_c(q-Q_0)}{N_L} H(q, \im
 \omega_n), \\
 A_\sigma (q, \omega_n) &= - \frac{v_s(q)}{N_L} \left\{ 1 + 4 v_s(q) K(q, \im
 \omega_n) \right\}, \\
 B_\sigma (q, \omega_n) &= - \frac{4 v_s(q) v_s(q-Q_0)}{N_L} H(q, \im
 \omega_n). 
\end{align} 


\end{document}